\documentclass[11pt]{article}
\usepackage{moriond,epsfig}

\def\be{\begin{equation}}
\def\ee{\end{equation}}
\def\bea{\begin{eqnarray}}
\def\eea{\end{eqnarray}}

\def\sun{\hbox{$\odot$}}
\def\micron{\hbox{$\mu$m}}
\def\degr{\hbox{$^\circ$}}
\def\arcsec{\hbox{$^{\prime\prime}$}}
\def\apj{ApJ}%
\def\aap{A\&A}%
\def\mnras{MNRAS}%
\begin{document}
\vspace*{4cm}
\title{DUST TEMPERATURES IN ASYMMETRIC PRESTELLAR CORES}

\author{D.~STAMATELLOS \& A.~P.~WHITWORTH}

\address{School of Physics \& Astronomy, Cardiff University, \\
          5 The Parade, Cardiff CF24 3YB, Wales, UK}

\maketitle\abstracts{
We present 2D Monte Carlo radiative transfer simulations of 
 flattened prestellar cores.
We argue the importance of observing prestellar cores 
near the peak of their emission spectra, 
and we point out observable characteristic 
features on isophotal maps of asymmetric cores at 
FIR wavelengths 
that are indicative of the cores' density and temperature structure.
These features are on scales 1/5 to 1/3 of the overall core size, and
so high resolution observations are needed to  observe them.
Finally, we discuss the importance of determining the temperatures of
prestellar cores with high accuracy.
}

\noindent
{\small¥{\it Keywords}: Stars: formation -- ISM: clouds-structure-dust -- 
Methods: numerical -- Radiative transfer}

\section{Introduction}

Prestellar cores are condensations in molecular clouds that are on the verge 
of collapse or already collapsing. Their study is important for constraining 
the initial conditions for star formation 
(see Andr\'e et al.~2000). 
Prestellar cores have typical sizes $2000-15000$~AU, and typical masses 
$0.05-10~{\rm M}_{\sun}$. They are cold with typical dust temperatures
$7-20$~K.
Their density profiles tend to be flat in the centre 
and to fall off as $r^{-n}$ ($n=2-4$) in the envelope. 
Many authors have used  Bonnor-Ebert (BE) spheres to represent prestellar
cores (e.g. Alves et al.~2001). However, studies
of the shapes of cores indicate that they are
best described as triaxial ellipsoids, and they are close to being
oblate spheroids (Goodwin et al.~2002).
Additionally, numerical simulations of the collapse of turbulent
molecular clouds indicate that prestellar cores are transient, non-spherical
systems, which when projected onto the plane of sky appear to mimic  BE spheres 
(Ballesteros-Paredes et al.~2004).
Information about prestellar cores comes from molecular line observations
(e.g. NH$_3$, CO), but also from continuum observations,
where cores are seen in emission (FIR, submm and mm) or absorption (NIR).
In this paper, we focus on modelling the continuum emission of prestellar 
cores.

\section{Asymmetric models of prestellar cores}

We use 2D models to 
represent slightly asymmetric 
prestellar cores (see Stamatellos et al.~2004). 
Our goal is to capture generically the different features we 
might hope to detect on the isophotal maps of such cores. 
We use  a  Plummer-like density profile (Plummer 1915),
that is  modified to include azimuthally 
symmetric departures from spherical symmetry.
This profile is simple and captures the basic observed characteristics
of prestellar cores.
Here, we present a model of a slightly flattened core 
(disk-like asymmetry; Fig.~\ref{fig.dens.temp}, left),
with density  profile
\begin{equation}
n(r,\theta) = n_0\,{\left[1 + A \left( \frac{r}{r_0} \right)^2 
{\rm sin}^p(\theta) \right]}{ \left[ 1 + \left( \frac{r}{r_0} \right)^2 
\right]^{-(\eta+2)/2} } \,,
\end{equation}
where $n_0$ is the density at the centre of the core, and 
$r_0$ is the extent of the central region in which the density is 
approximately uniform. 
The parameter $A$ determines the equatorial-to-polar optical 
depth ratio $e\,$, i.e. the maximum optical depth from the centre 
to the surface of the core ($\theta = 90\degr$), 
divided by the minimum optical depth from the centre 
to the surface of the core ($\theta = 0\degr$ 
and $\theta = 180\degr$). The parameter $p$ determines how rapidly 
the optical depth from the centre to the surface rises with 
increasing $\theta$, i.e. going from the north pole at $\theta 
= 0\degr$ to the equator at $\theta = 90\degr$.
We assume that the core has a spherical boundary at radius 
$R_{\rm core} = 2 \times 10^4\,{\rm AU}$, and $n_0 = 10^6\,
{\rm cm}^{-3}$, $r_0 = 2 \times 10^3\,{\rm AU}$, $\eta = 2$, $p=4$, and
 $e=2.5$. The simulations are performed using {\sc Phaethon}, a  3D Monte Carlo
radiative transfer code (Stamatellos  \& Whitworth 2003).

\section{Dust temperature profiles}

\begin{figure}[hb]
\centerline{
\includegraphics[width=6cm,angle=-90]{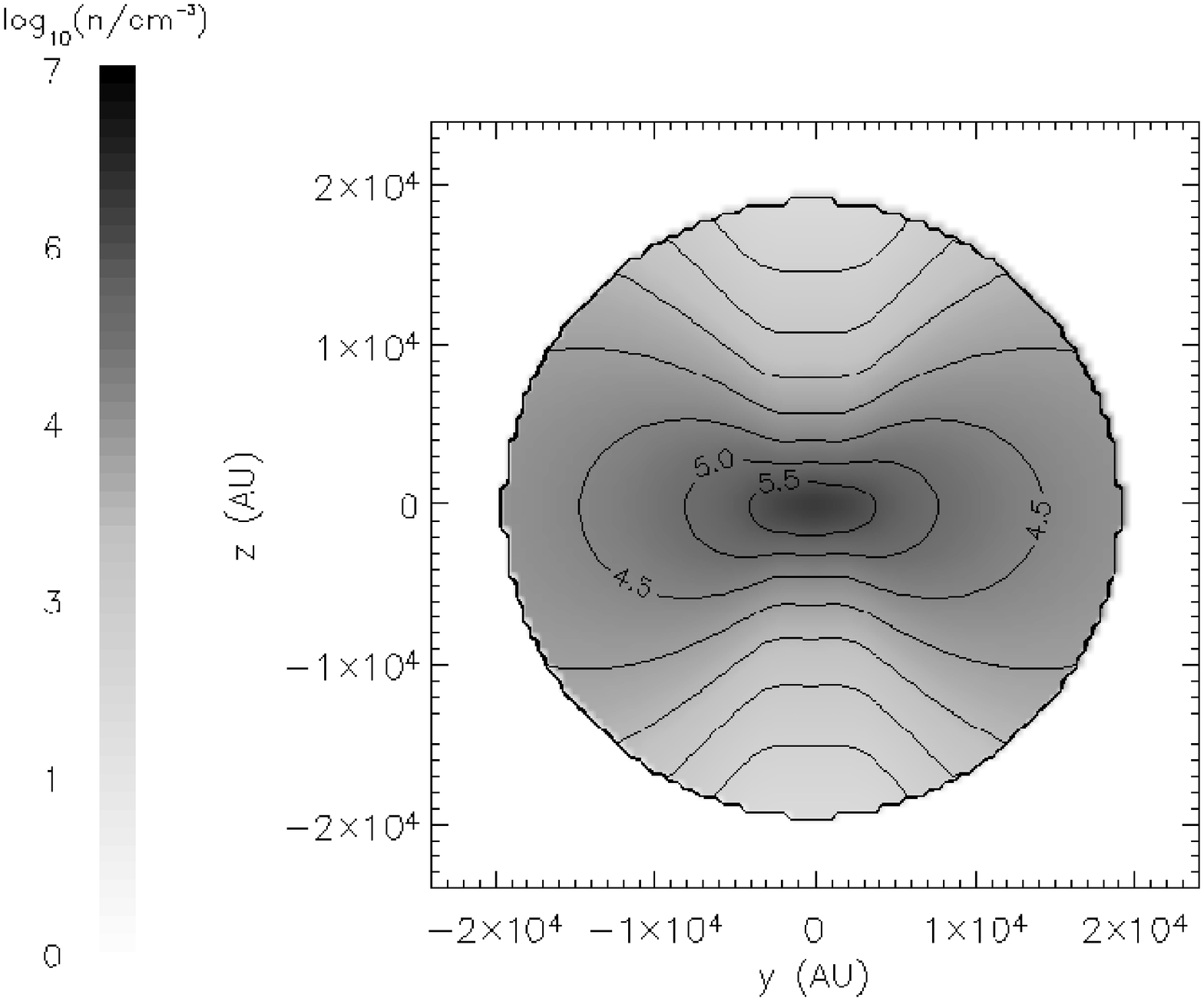}\hspace{1em}
\includegraphics[width=6cm,angle=-90]{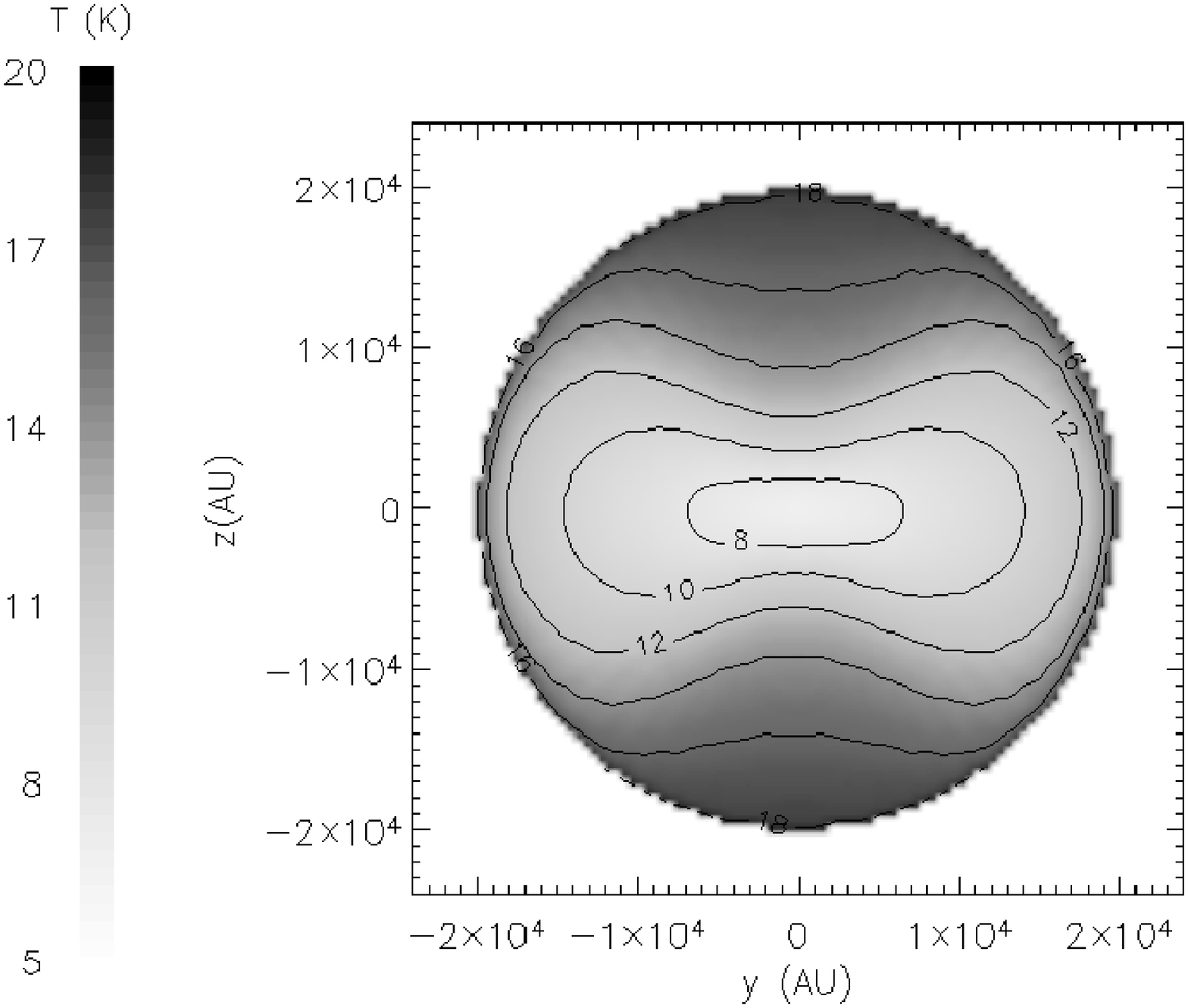}}
\caption{{\bf Left:} Density on the $x=0$ plane for a flattened 
asymmetric core with equatorial-to-polar optical depth ratio $e=2.5$ 
and $p=4$. We plot isopycnic contours every $10^{0.5} {\rm cm^{-3}}$. 
The central contour corresponds to $n=10^{5.5} {\rm cm^{-3}}$.
{\bf Right:} Temperature on the $x=0$ plane for the same model,
calculated with a Monte Carlo RT simulation.
We plot isothermal contours from 8 to 18~K, every 2~K.}
\label{fig.dens.temp}
\label{fig_temp.asyma}
\end{figure}

The dust temperature (Fig.~\ref{fig_temp.asyma}, right) drops from 
around 17~K  at the edge of the core to 7~K at the centre of the core
(also see Zucconi et al. 2001; Evans et al. 2001; 
Stamatellos \& Whitworth 2003). 
We  find that the dust temperature 
inside a core with disk-like asymmetry is $\theta$ dependent, 
similar to the results of Zucconi et al.~(2001) and 
Gon\c{c}alves et al.~(2004). 
As expected, the equator of the core is colder than the poles. 
The difference in temperature between two points having the same distance 
$r$ from the centre of the core but with different polar angles $\theta$, 
is larger for  more asymmetric cores (see Stamatellos et al.~2004). 
This temperature difference will affect the appearance of the core at 
wavelengths shorter than or near the core peak emission.

\section{SEDs and isophotal maps}

\begin{figure}[ht]
\centerline{
\includegraphics[width=6.cm,angle=-90]{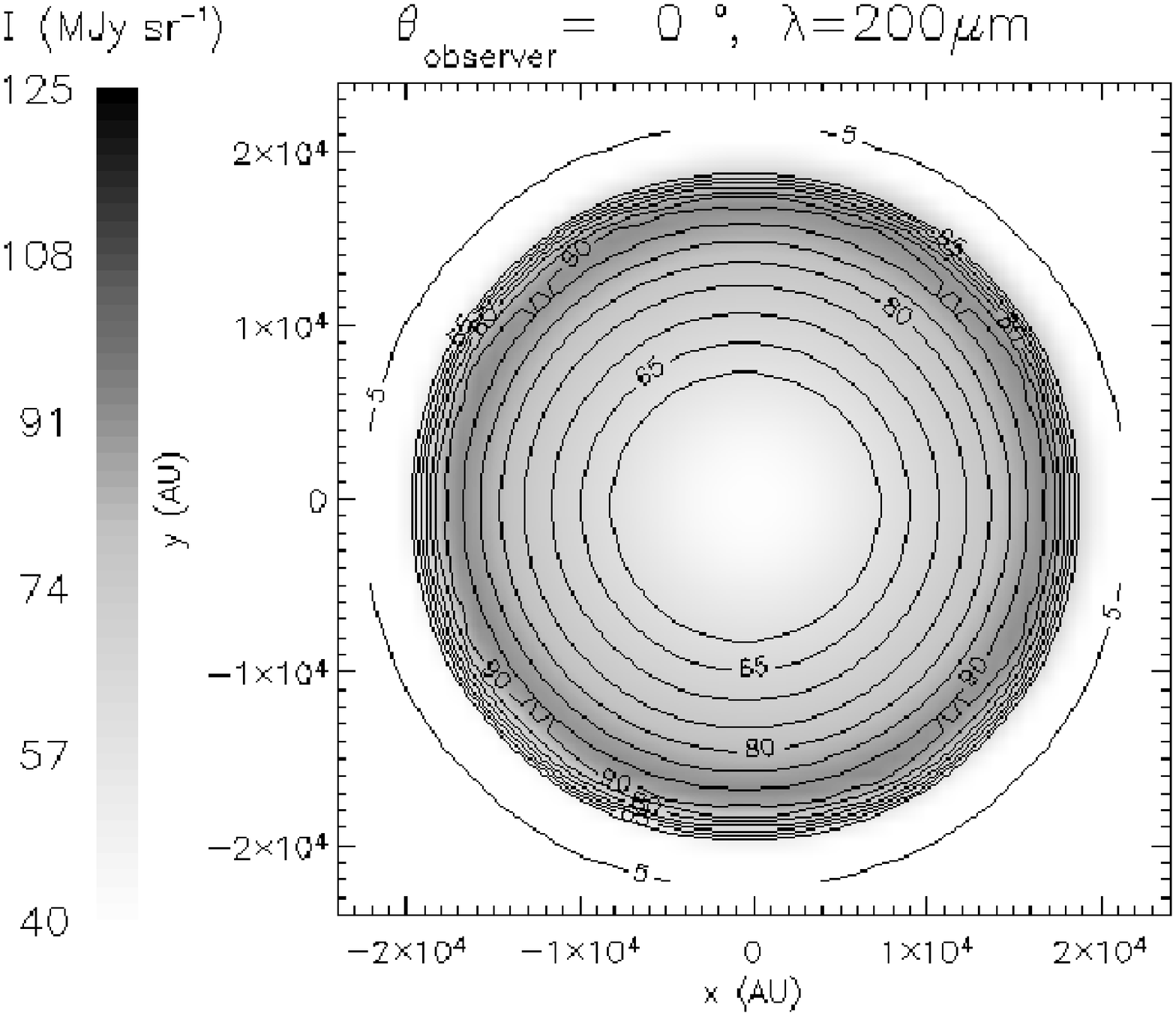}\hspace{1cm}
\includegraphics[width=6.cm,angle=-90]{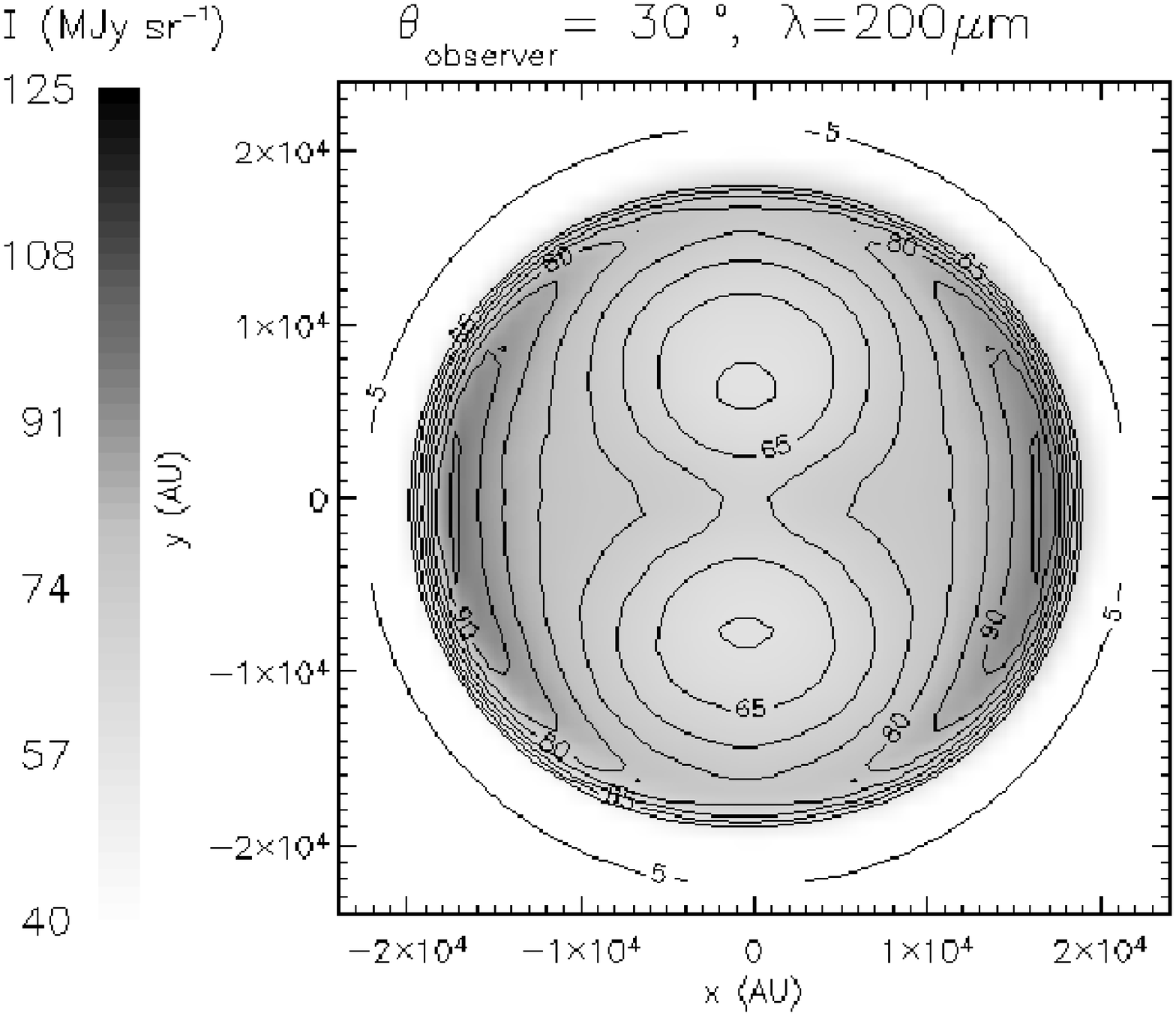}}
\centerline{
\includegraphics[width=6.cm,angle=-90]{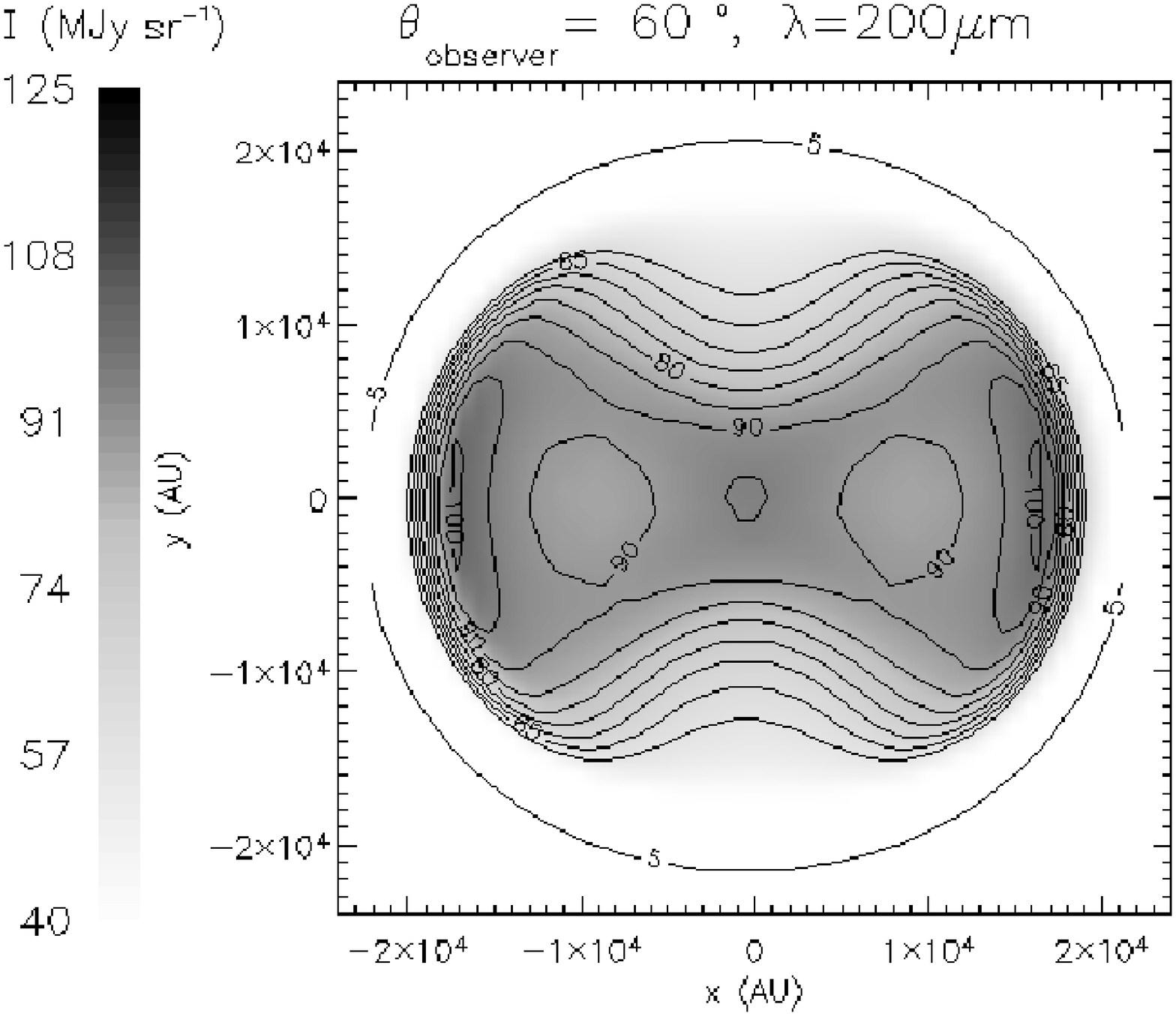}\hspace{1cm}
\includegraphics[width=6.cm,angle=-90]{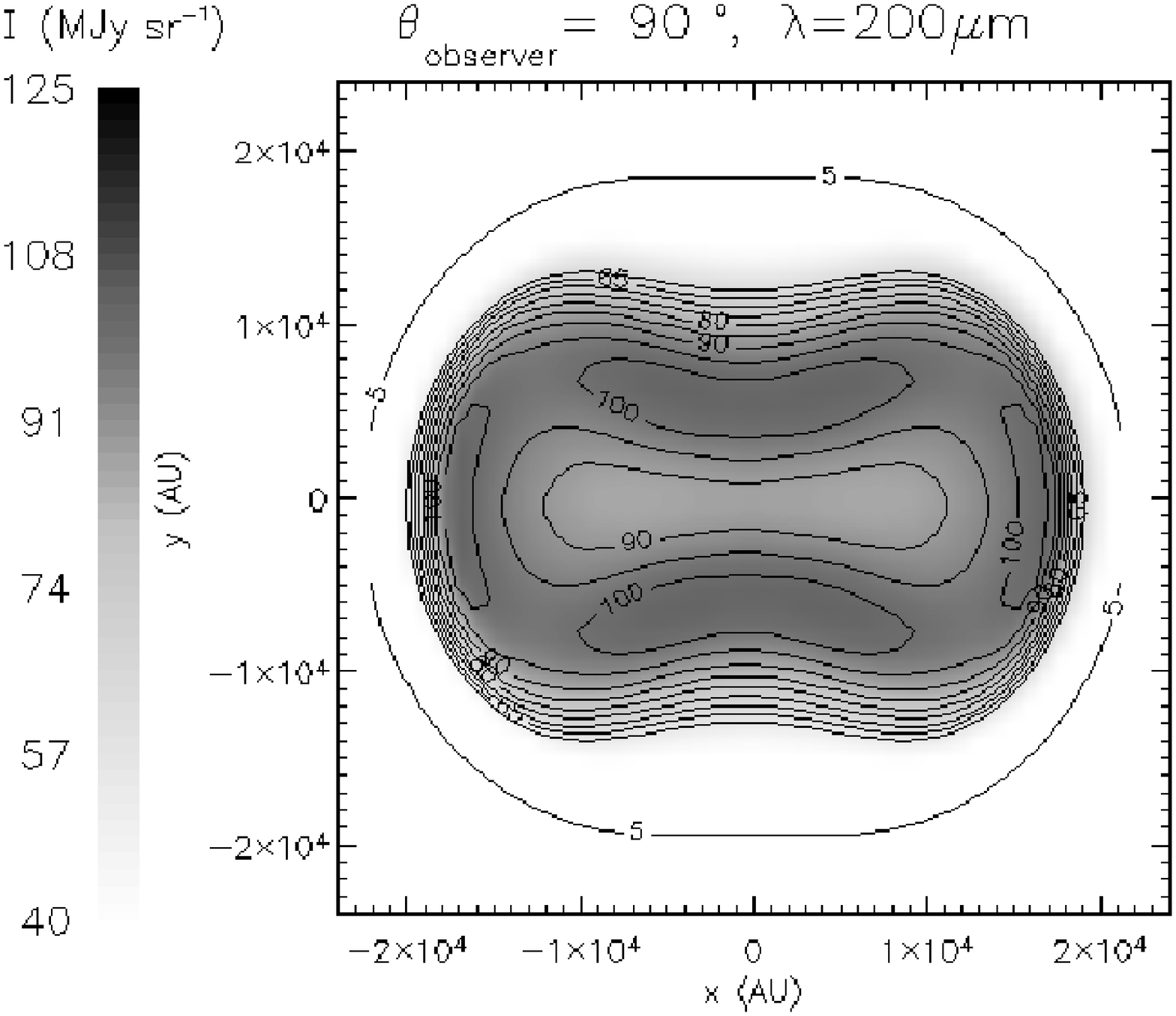}}
\caption{Isophotal maps at viewing angles $0\degr$, $30\degr$
$60\degr$ and $90\degr$,  for a flattened prestellar core with 
$e=2.5$  and $p=4$ at 200~$\micron$.
We plot an isophotal contour at 5~MJy~sr$^{-1}$ and then from 60 to
110~MJy~sr$^{-1}$, every 5~MJy~sr$^{-1}$. There are characteristic 
symmetric features due to core temperature and orientation with respect 
to the observer. 
(We note the axes $(x,y)$ refer to the plane of sky 
as seen by the observer).
} 
\label{fig.iso}
\label{images.asyma.2.5}
\end{figure}

The SED of the slightly asymmetric core
we examine, is the same at any viewing angle, because 
the core is optically thin to the radiation it emits (FIR and longer 
wavelengths). 
However, the isophotal maps do depend on 
the observer's viewing angle. 
Additionally, they depend on the wavelength of observation. 
{\sc Phaethon} calculates images at any wavelength, and therefore provides a 
useful tool for direct comparison with observations, e.g.  at MIR 
(ISOCAM), FIR (ISOPHOT) and submm/mm (SCUBA, IRAM) 
wavelengths. We focus on
wavelengths near the peak of the core emission (150-250$~\micron$; 
we choose 200$~\micron$ as a representative wavelength).
At 200~$\micron$ the core appearance depends both on its temperature and its 
column density in the observer's direction.  
This interplay between core temperature  
and column density along the line of sight results in characteristic 
features on the images of the cores (see Fig.~\ref{fig.iso}). 
Such features include (i) the two 
intensity minima  at almost symmetric positions relative to the centre of 
the core, on the image at 30\degr, and (ii) the two 
intensity maxima, again at  symmetric positions relative to the centre 
of the core, on the image at 90\degr.

\begin{figure}
\centerline{
\includegraphics[width=5.25cm]{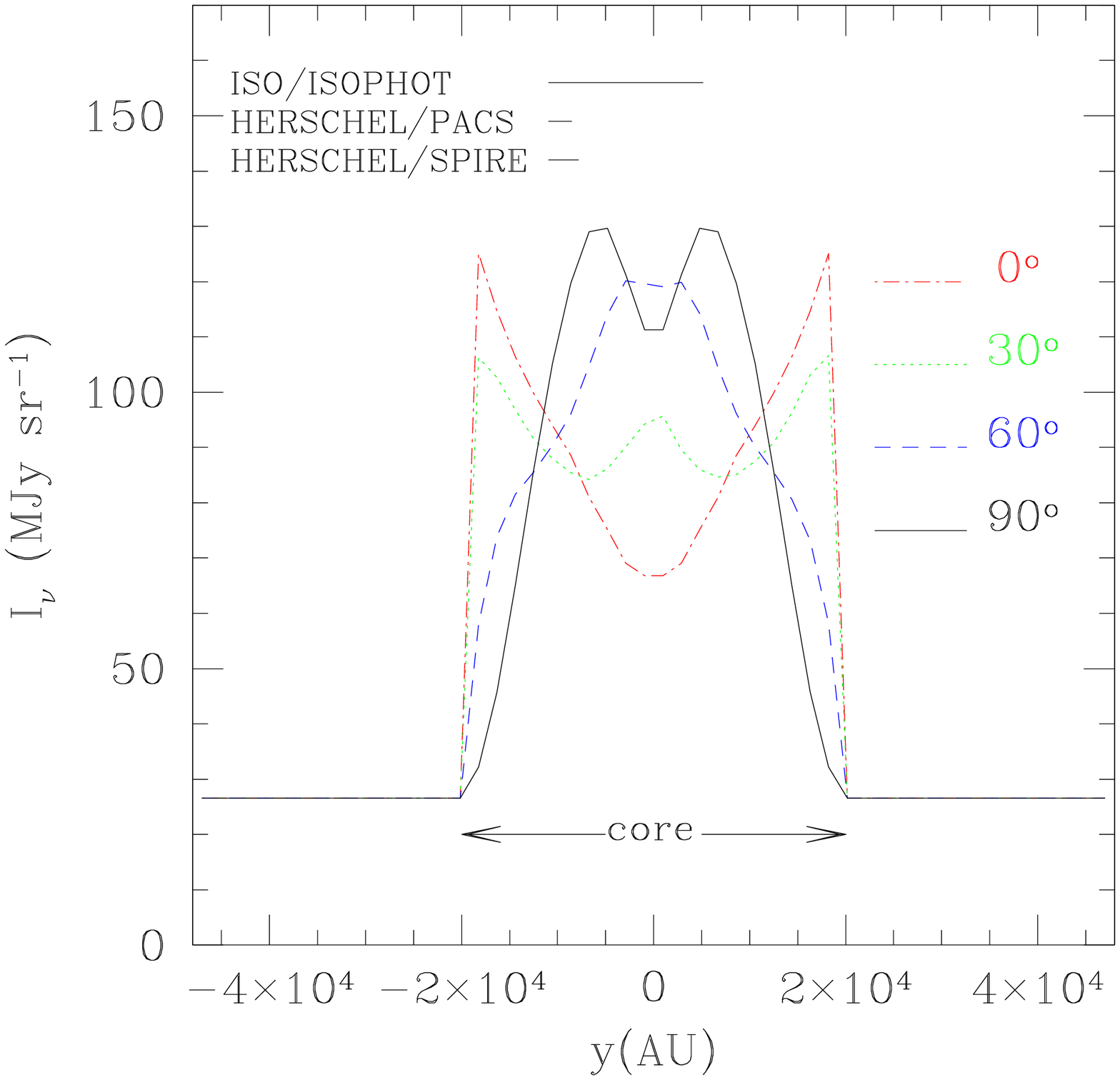}
\includegraphics[width=5.45cm]{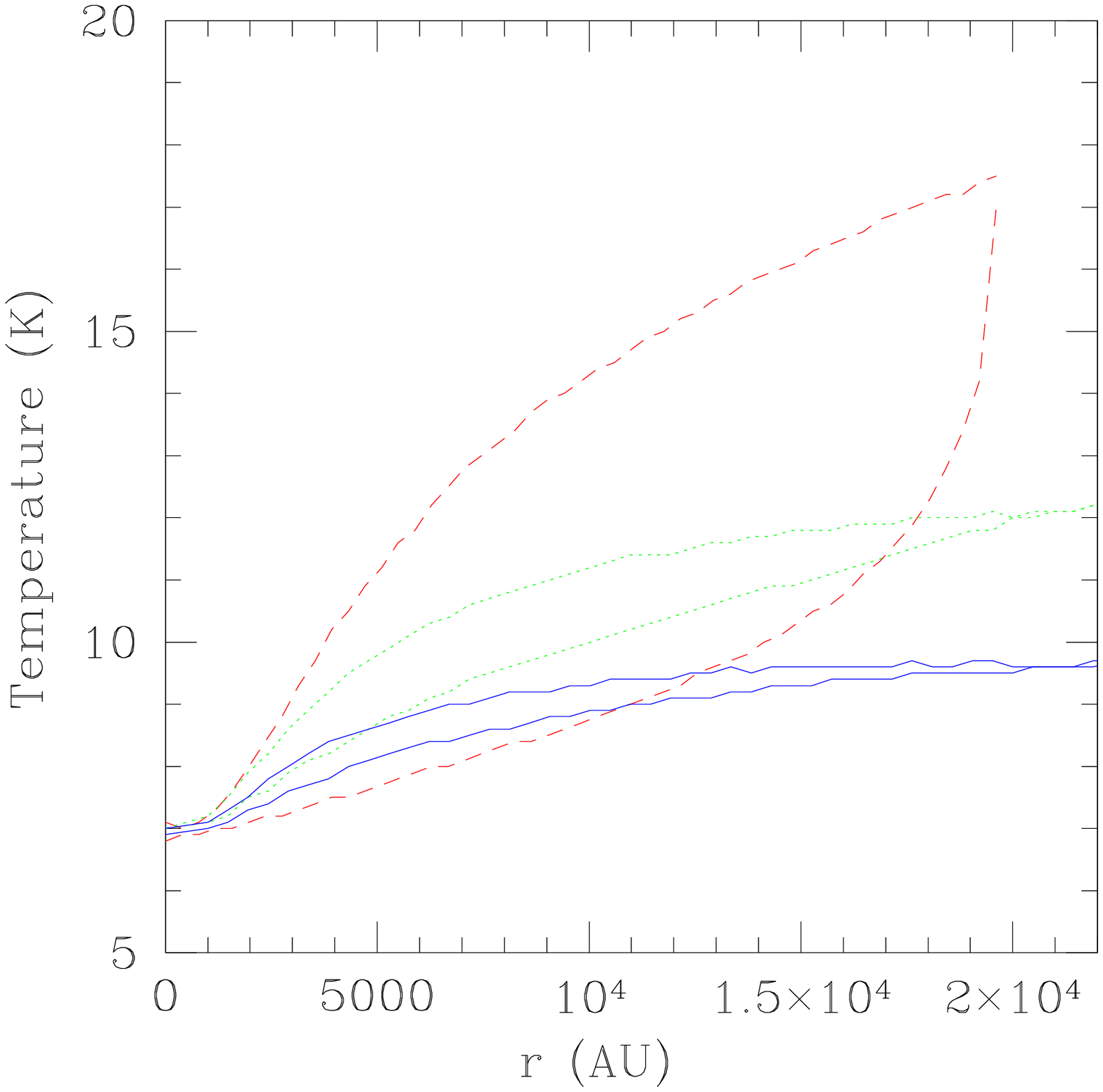}
\includegraphics[width=5.25cm]{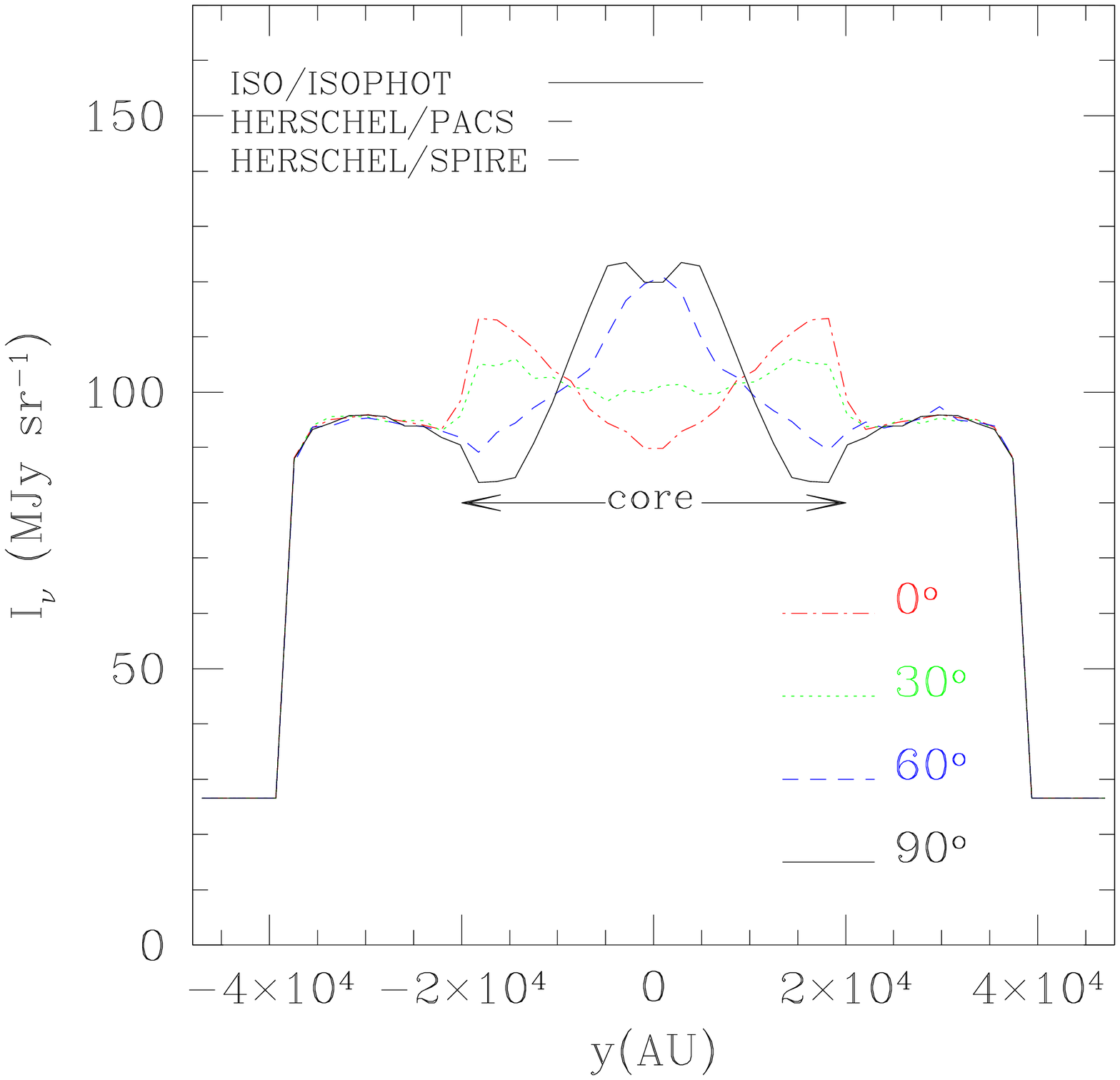}}
\caption{{\bf Left:} A perpendicular cut through the centre of the 
core images presented in Fig.~\ref{images.asyma.2.5} 
(but also including the background radiation field). We also
plot the beam size of ISOPHOT 
 and the beam size of the upcoming {\it Herschel} PACS/SPIRE
(assuming a core distance of 100~pc). 
{\bf Centre:} The effect of the parent cloud on cores. Temperature profiles 
of a non-embedded core (dashed lines), and of a core at the 
centre of an ambient cloud with  $A_{\rm V}=4$ (dotted 
lines), and $A_{\rm V}=13$ (solid lines). The upper curve of each set of 
lines corresponds to the direction towards the pole of the core 
($\theta=0\degr$), and the bottom curve to the direction towards the 
core equator ($\theta=90\degr$). The difference between the two curves
is indicative of the temperature gradient.
The core is colder when it is 
inside a thicker parent cloud, and the temperature gradient within the core 
is smaller.
{\bf Right:} Same as on the left, but for a 
core embedded in a molecular cloud with visual extinction 
$A_{\rm V}=4$. The characteristic features at different viewing 
angles are weaker than in the case of a non-embedded core,
 but still observable.}
\label{image_rprof_nem}\label{image_rprof_em}
\end{figure}

We conclude that  isophotal maps at 200~$\micron$ contain detailed  
information, and sensitive, high resolution observations at 
200~$\micron$ are helpful in constraining the core density 
and temperature structure and  the orientation of the core with 
respect to the observer. In Fig.~\ref{image_rprof_nem}, left, we present 
a perpendicular cut (at $x=0$) of the core images shown in 
Fig.~\ref{images.asyma.2.5}. We also plot the beam size of the 
ISOPHOT C-200 camera (90\arcsec, or 9000~AU for a core at 100 pc) 
and the beam size of the upcoming (2007) {\it Herschel} (13\arcsec 
or 1300~AU for the $170~\micron$ band of PACS; 17\arcsec or 1700~AU 
for the $250~\micron$ band of SPIRE). ISOPHOT's resolution is probably 
not good enough to detect the features mentioned above. Indeed, a 
search in the Kirk~(2002) sample of ISOPHOT observations 
(also see Ward-Thompson et al.~2002) does not reveal any cores with 
such distinctive features. However, {\it Herschel} should, in 
principle, be able to detect such features in the future.

\section{The effect of the parent cloud}

Cores are generally embedded in molecular clouds, with visual optical 
depths ranging from 2-10 (e.g. in Taurus) up to 40 (e.g. in $\rho$ 
Oph). Due to the  ambient cloud  the radiation 
incident on a core embedded in the cloud is reduced in the UV and 
optical, and enhanced in  the FIR and submm (Mathis et al. 1983). Previous 
radiative transfer calculations of spherical cores embedded at the centre 
of an ambient cloud (Stamatellos \& Whitworth 2003), have shown that 
embedded cores are colder ($T<12$~K) and that the temperature gradients 
inside these cores are smaller than in non-embedded cores. 

Here, we examine the same core as before when it is embedded
in a uniform density ambient cloud with different visual 
extinctions $A_{\rm V}$. The ambient cloud is illuminated by the interstellar
radiation field.
Relative to the non-embedded core, the core embedded in an ambient cloud 
with $A_{\rm V} = 4$ is colder and has lower temperature gradient 
(Fig.~\ref{image_rprof_em}, centre). 
The isophotal maps are similar to those of the 
non-embedded core, but the characteristic features 
are less pronounced. This is because the temperature gradient inside 
the core is smaller when the core is embedded. 
In Fig.~\ref{image_rprof_em}, right, 
we present a perpendicular cut through the centre of the embedded core 
image. It is evident that the features are 
quite weak, but they have the same size as in the non-embedded core 
(Fig.~\ref{image_rprof_nem}, left), and they should be detectable with {\it 
Herschel}, given an estimated rms sensitivity better than $\sim 1-3$ 
MJy sr$^{-1}$ at 170-250~$\micron$.

Thus, continuum observations near the peak of the core emission can 
be used to obtain information about the core density and temperature 
structure and orientation, even  if the core is very embedded 
($A_{\rm V}\sim 10$).

\section{Application: Modelling L1544}

\begin{figure}[hb]
\centerline{
\includegraphics[height=6.cm]{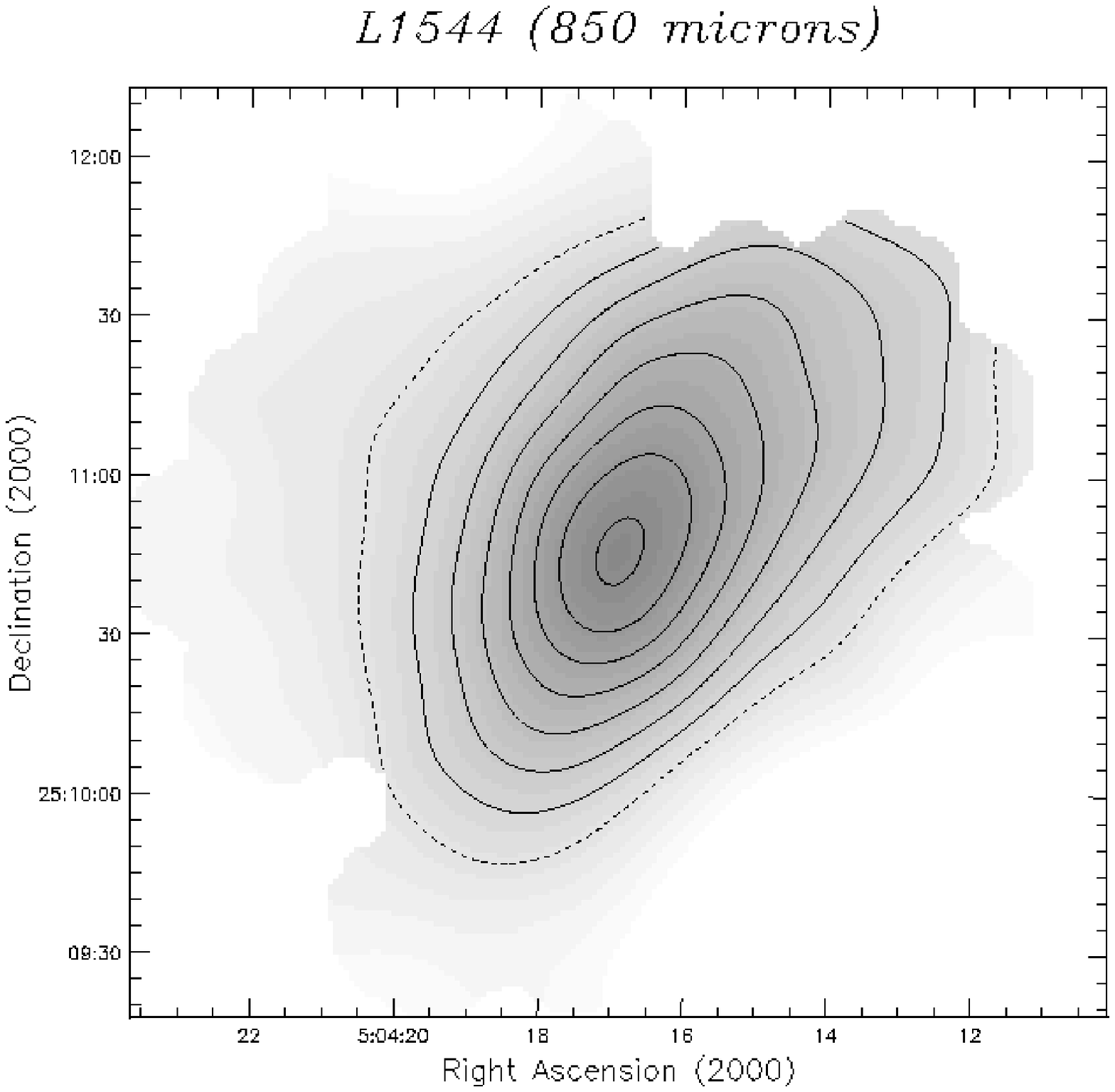}\hspace{1.5cm}
\includegraphics[width=6.5cm,height=6.2cm]{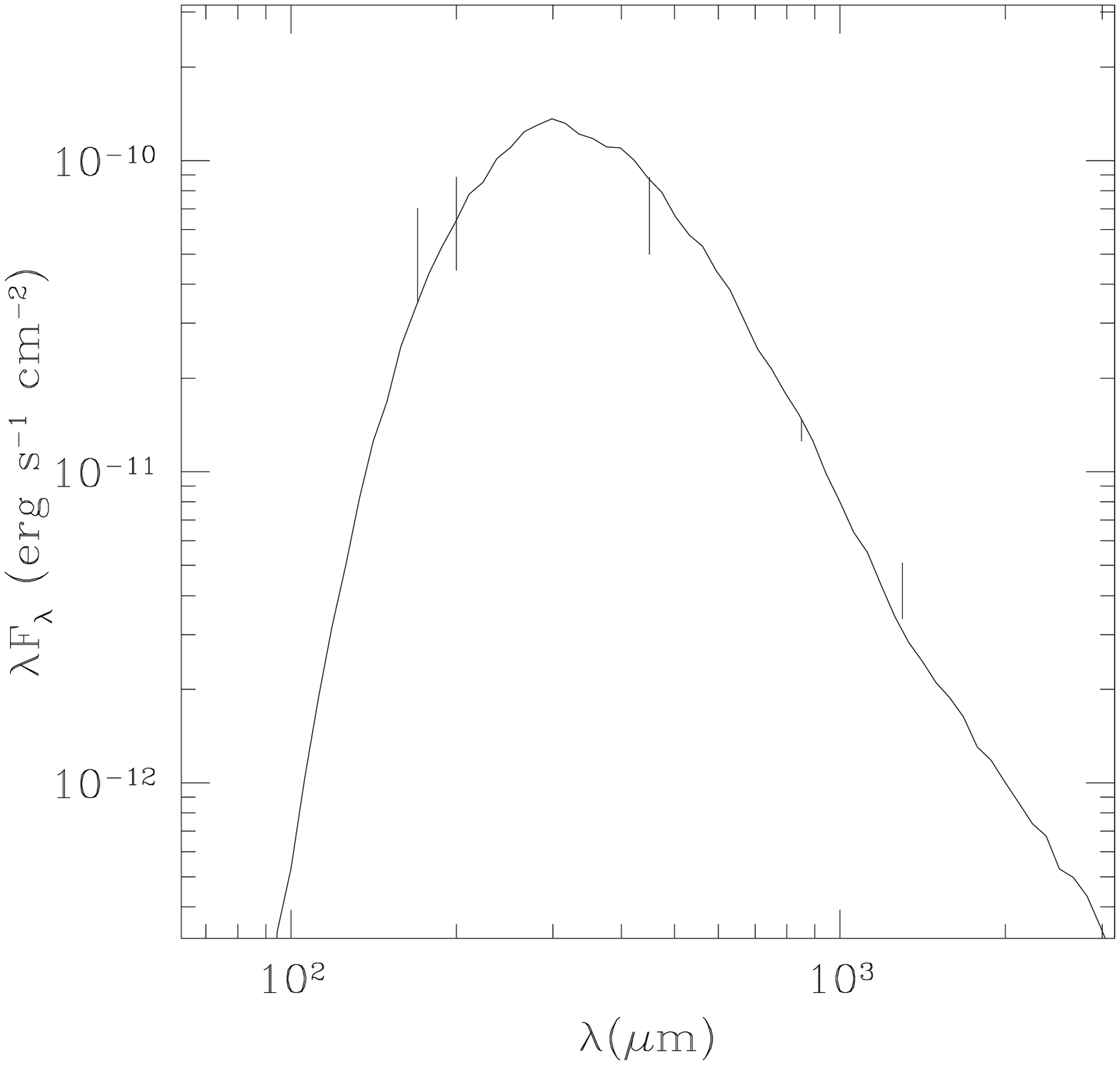}}
\caption{
{\bf Left:} 850~$\micron$ SCUBA image of L1544 (from Kirk 2002).
{\bf Right:} SED of the core. The line corresponds to the model, and the
points to the observed SED (data taken from Kirk 2002). 
}
\label{fig_obs}  
\centerline{
\includegraphics[height=6.4cm]{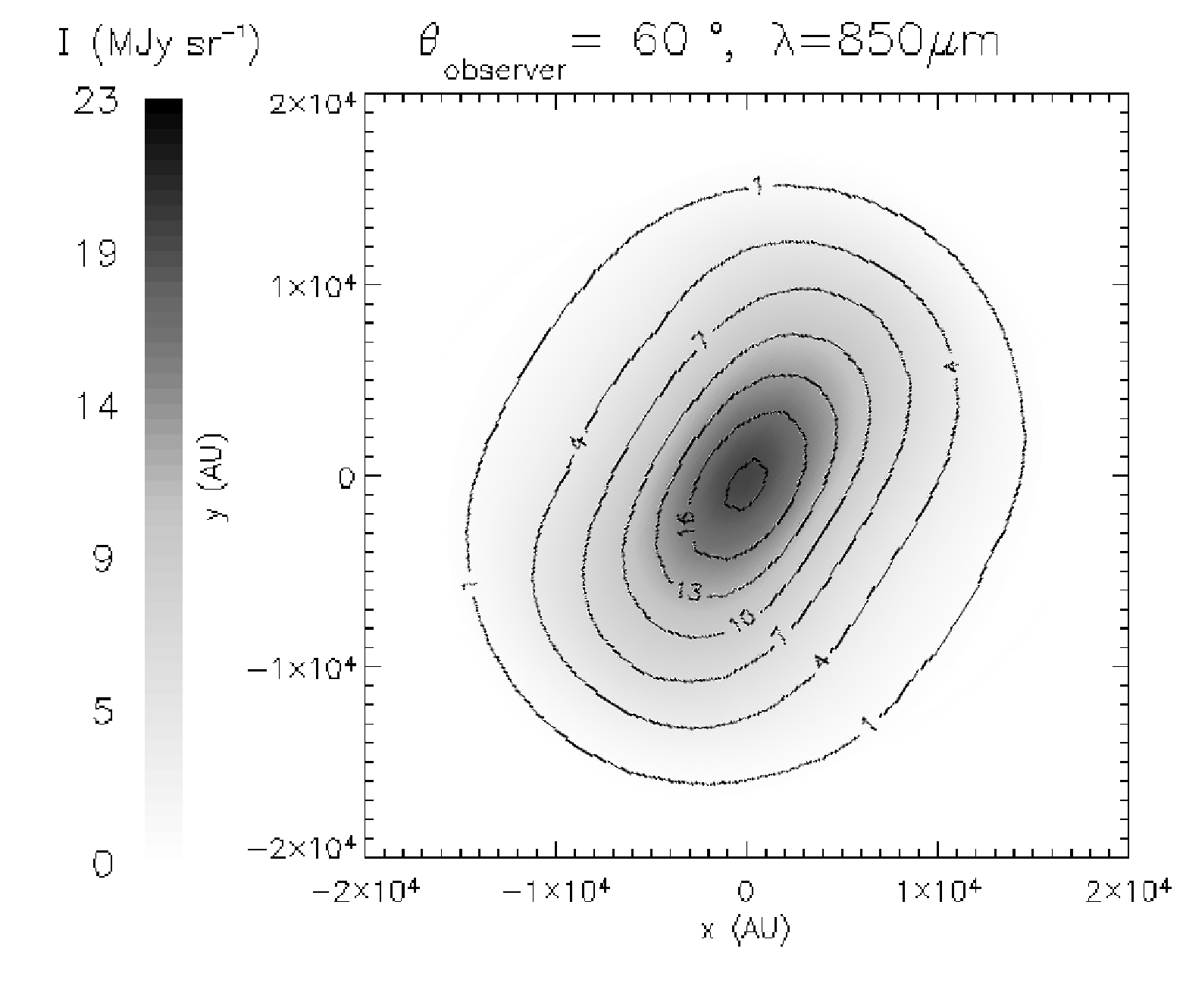}\hspace{0.5cm}
\includegraphics[width=6.5cm]{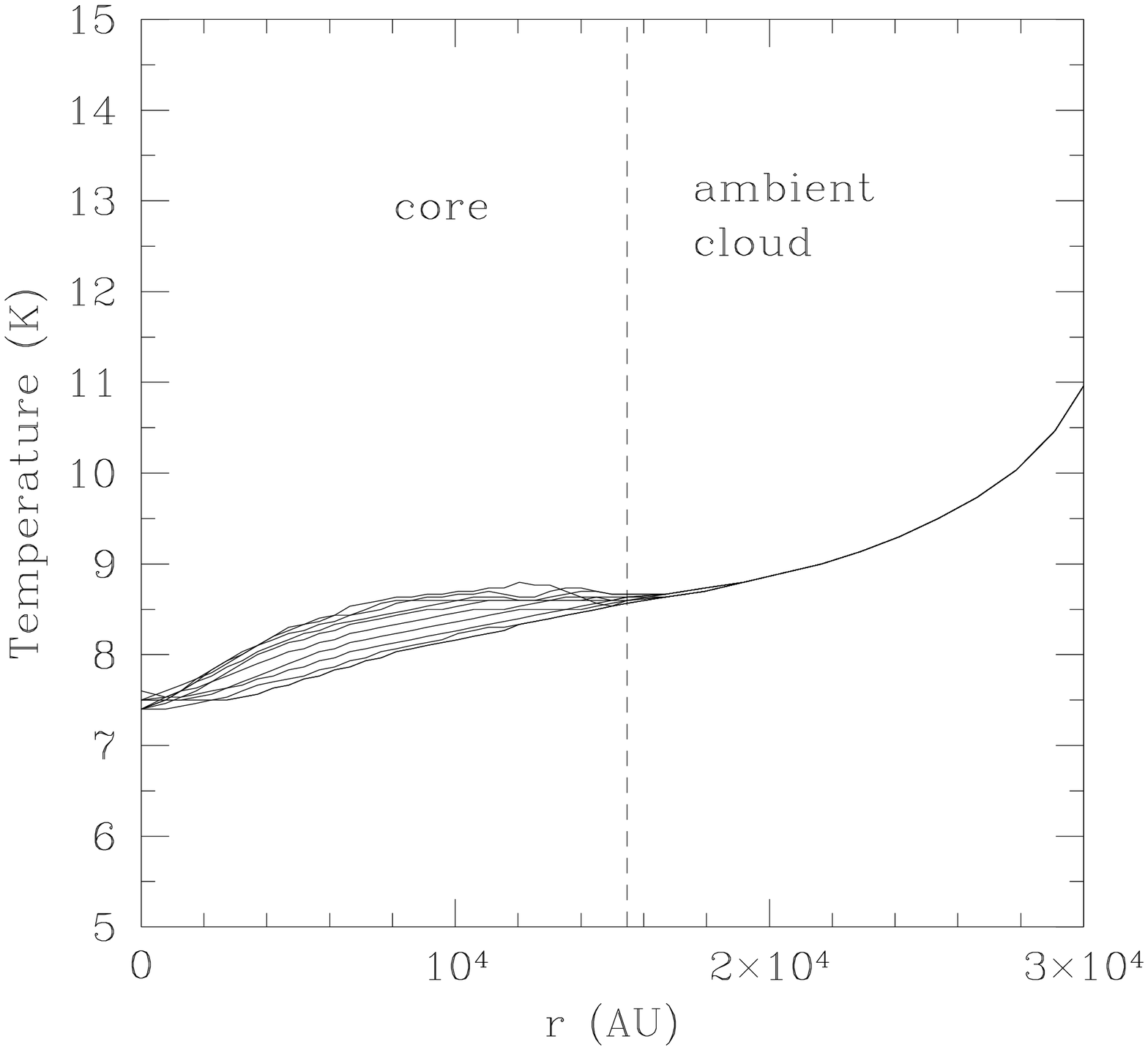}}
\caption{
{\bf Left:} 850~$\micron$ isophotal map of the model. 
The core is viewed at an angle $\theta=60\degr$. 
{\bf Right:} Dust temperature profile of the core and the immediate 
ambient cloud.
Different lines correspond to different directions 
(from $\theta=0\degr$, upper line, to  $\theta=90\degr$, bottom line).
}
\label{fig_model}
\end{figure}

We model L1544 (core distance
$D=140~{\rm pc}$) using a slightly flattened density profile
($e=2$ and $p=4$).
Our goal is to fit  the SCUBA 850~$\micron$ image 
(Fig.~\ref{fig_obs}, left), 
the SED data points (Fig.~\ref{fig_obs}, right),
and to calculate the temperature profile in the core.
To fit the observational data we choose the
asymmetry parameter $e$, based on the 850~$\micron$ image of the core.
Then we vary the mass of the core (by adjusting the central core density
and the core flattening radius), so as to fit the submm and mm SED data.
Finally, we vary the extinction through the ambient cloud in order to
fit the FIR SED data.

The best-fit model is obtained using central density
$n_0 = 3.4\times10^5\,{\rm cm}^{-3}$,
inner flattening radius $r_0 =0.015\,{\rm pc}$,
core extent $R_{\rm core} = 0.075\,{\rm pc}$, and an ambient cloud 
of visual extinction
$A_V^{\rm cloud}=23.7$. These values are in good agreement with observations of
L1544 (e.g. Ward-Thompson et al.~2002, Kirk et al.~2004).
The calculated luminosity emitted from the core is 
$L_{\rm core}=0.10$~L$_{\sun}$, in agreement with observations (Kirk 2002).
The  model fits the SED data very well (Fig.~\ref{fig_obs}, right). 
The model also reproduces reasonably well  the central region of the L1544
850~$\micron$ map (cf. Figs~\ref{fig_obs} and \ref{fig_model}, left).
However, it does not reproduce the asymmetries in the outer parts
of the core, which are signatures of triaxiality.

The temperature at the
edge of the core is
$T_{\rm edge}=8.7$~K
and at the centre $T_{\rm centre}=7.5$~K 
(Fig.~\ref{fig_model}, left). The dust temperature we calculate with the 
model is lower by  $2-3$~K than the 
temperature estimated by Kirk (2002) using FIR (90, 170 and 200~$\micron$) 
ISOPHOT observations ($T_{\rm iso}=10.2^{+0.5}_{-0.4}$~K).
However, ISO observations have difficulty  distinguishing the core from
the ambient cloud, and thus this larger temperature may be due to 
the presence of the hotter ambient cloud in the observing beam.
As a result of overestimating the dust temperature, the core mass calculated
by Kirk (2002) is underestimated. 
The core mass is 
$M_{\rm core} \propto F_\lambda/
\left(B_\lambda (T_{\rm dust})\kappa_\lambda\right)$
(e.g. Andr\'e et al.~1999),
where $F_\lambda$ is the flux at a submm or mm wavelength, and
${\kappa_\lambda}$ is the assumed dust opacity per unit mass.
If the core temperature is overestimated 
by 2~K (10~K instead of 8~K) then using the 850~$\micron$ flux
the core mass is underestimated by a factor of 1.6
($B_{\rm 850\micron}(10{\rm K})= 1.6\times B_{850\micron}\rm (8{\rm K})$).
The core mass of the model is
$M=2.5$~M$_{\sun}$ which is similar to what Kirk et al.~(2002) calculated
($M_{\rm iso}^{\rm  850\mu {\rm m}}=2.5 \pm 1$).
However, we note that  Kirk  uses a 
dust opacity  $\kappa_{850\mu m}=0.01~{\rm cm}^2 {\rm g}^{-1}$, which is 
smaller by a factor of 2 than the dust opacity we use 
($\kappa_{850\mu m}=0.02~{\rm cm}^2 {\rm g}^{-1}$). 
Thus, to consistently compare the mass 
we need to divide the Kirk (2002) value by 2.

We conclude that accurate estimates of core 
temperatures are important when calculating core masses from submm and mm
observations, since e.g.  
overestimating temperatures by even just  1-2 K can lead to underestimating 
core masses by a factor of ~2. 

\section{Conclusions}
 
Far-infrared continuum maps of prestellar cores reflect both the column 
density and  temperature field along the line of sight, and thus 
contain complementary information to the mm continuum maps that
mainly trace column density. The radiative transfer models presented
here show that the effect of the combined dust temperature and
column density along the line of sight is to produce characteristic
features in the FIR intensity maps. These features are useful for
constraining the conditions in prestellar cores, and 
are expected to be present in cores with
high enough  temperature gradients ($\sim$2~K).

\section*{Acknowledgements}

We  acknowledge help from the EC Research Training Network
``The Formation and Evolution of Young Stellar Clusters'' (HPRN-CT-2000-00155).
We also thank  P.~Andr\'e for useful comments.

\end{document}